\documentclass{webofc}

\usepackage[varg]{txfonts}   
\usepackage{hyperref}
\usepackage{url}

\hypersetup{colorlinks=true,citecolor=blue,urlcolor=blue,linkcolor=blue}
\begin{document}
\newcommand{\lambdac}{$\Lambda_c^+$\xspace}
\newcommand{\dzero}{$D^0$\xspace}
\newcommand{\pp}{$p$$+$$p$\xspace}
\title{Towards a precise measurement of the \lambdac/\dzero ratio at RHIC}
%
%

\author{{\firstname{Joseph D.} \lastname{Osborn}\inst{1}\fnsep\thanks{\email{josborn1@bnl.gov}}for the sPHENIX Collaboration
}}

\institute{Brookhaven National Laboratory
          }

\abstract{
sPHENIX is a next-generation experiment at RHIC for jet and heavy-flavor physics which was fully commissioned during 2023 and 2024. Using its novel streaming-readout-capable, precision tracking system, sPHENIX collected 100 billion unbiased \pp collisions, and a further sample of minimum-bias Au-Au collisions, in Run-24. A key measurement of the sPHENIX heavy flavor physics program is the comparison of \lambdac to \dzero differential yields in both Au+Au and p+p collisions, which probes questions related to the hadronization of heavy-flavor baryons compared to mesons in the Quark-Gluon Plasma medium and in vacuum. At RHIC energies, there is no previous measurement of the \lambdac/\dzero baseline in \pp collisions, modern Monte Carlo event generators give widely different predictions, and the ratio in Au+Au is only poorly known. These proceedings present the status of measurement from sPHENIX of the \lambdac/\dzero ratio in p+p collisions.
}
\maketitle
\section{Introduction}
\label{intro}
The sPHENIX experiment is a brand new state-of-the-art collider physics detector built at the Relativistic Heavy Ion Collider (RHIC) at Brookhaven National Laboratory~\cite{PHENIX:2015siv}. sPHENIX is a large-acceptance, high-rate detector that has been optimized for the study of high $p_T$ probes of the quark gluon plasma (QGP) in heavy ion collisions. The primary goal of the experiment is to probe the QGP at shorter length scales by measuring jet and heavy flavor observables in a $p_T$ regime accessible at RHIC complementary to those at the Large Hadron Collider (LHC). In addition to this, sPHENIX has a wide breadth of physics interests studying QCD in small and large systems.

One of the primary physics motivations for sPHENIX is to study open heavy flavor particle production in both \pp and Au+Au collisions at RHIC. Due to their large masses, heavy quarks are produced in hard scattering processes early in the collision history and thus experience the full evolution of the QGP medium and of the hadronization process. The sPHENIX detector was designed to make precision measurements of heavy flavor charm and bottom quark hadrons. In particular, measuring the \lambdac/\dzero ratio in \pp collisions at RHIC is one of the primary goals of the experiment for providing new insight into hadronization mechanisms. This measurement has been of great interest due to large differences seen in $e^++e^-$ collisions~\cite{CLEO:2004enr}, collision systems involving hadrons (see e.g. Ref.~\cite{ALICE:2017thy,ALICE:2018hbc}), and Monte-Carlo event generators, suggesting additional hadronization mechanisms that remain to be fully understood. This motivates a measurement from sPHENIX, which will provide additional data at center-of-mass energies in between $e^++e^-$ colliders and the LHC.

sPHENIX was constructed in the years 2020-2022 and was commissioned using Au+Au beams in 2023 and \pp collisions in 2024 at $\sqrt{s}=200$ GeV. After commissioning, sPHENIX collected a large sample of unbiased \pp collisions using its unique streaming readout capabilities, described further in Sec.~\ref{sec-1}. These proceedings report the status of the track reconstruction commissioning and ongoing heavy flavor physics analysis at sPHENIX, looking towards the first measurement of the \lambdac/\dzero ratio measurement at RHIC. 

\begin{figure}
\centering
\includegraphics[width=0.8\linewidth,clip]{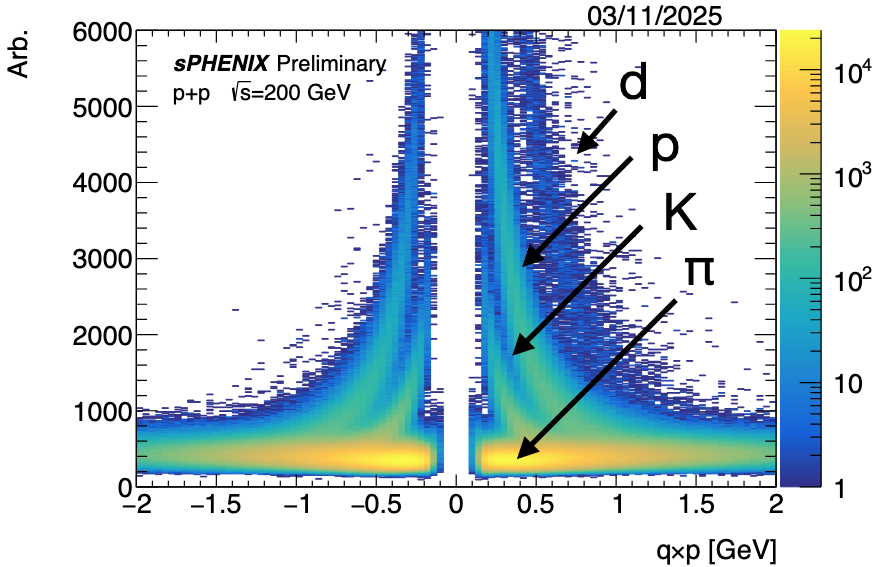}
\caption{ The $dE/dx$ distribution as a function of particle momentum times charge, showing the $dE/dx$ performance of the TPC with initial calibrations.}
\label{fig:dedx}       
\end{figure}

\section{sPHENIX Detector and Reconstruction}
\label{sec-1}

The sPHENIX experiment is comprised of precision tracking and calorimetery detectors covering a pseudorapidity acceptance of $|\eta|<1.1$ and full azimuth. There are four tracking subsystems used for reconstruction and subsequent heavy flavor resonance analysis. The MVTX is a precision vertex pixel detector that is constructed of 3 layers of MAPS staves within a few centimeters of the beam pipe in radius. The MVTX provides a track pointing resolution of approximately 30 microns; this is essential for separating short lived secondary decays from the primary vertex, a key aspect for heavy flavor reconstruction. The INTT is a silicon strip detector that provides additional precise measurements for pattern recognition. Most importantly, it has an integration time that is $\sim$100 ns, allowing tracks to be associated to particular bunch crossings from RHIC which delivers \pp collisions with 106.5 ns bunch spacings. The TPC is a compact GEM readout detector with 48 layers and provides the necessary momentum resolution for the sPHENIX tracking physics program. Figure~\ref{fig:dedx} shows the $dE/dx$ performance with the first set of calibrations of the TPC. Despite not initially being designed for PID, there is excellent separation power at low $p_T$ for heavy flavor combinatoric background suppression. Lastly, the TPOT is constructed of 8 tiles of micromegas detectors that are primarily used for calibrating the space charge distortions present in the TPC. All four tracking detectors provide critical information for four-dimensional reconstruction to precisely determine the time and space structure of physics events.

The tracking detectors can operate in a hybrid streaming readout mode of operation where some fraction of the total delivered luminosity is read out following a hardware trigger. This fraction is limited only by the total data volume readout from the TPC to storage, where the limitation comes from the available storage. The streaming mode of operation is motivated by the open heavy flavor physics program, where low~$p_T$~open heavy flavor hadrons have poor hardware trigger efficiencies. Concluding in October 2024, the tracking detectors at sPHENIX recorded unbiased \pp collisions at a rate of $\mathcal{O}(200)$~kHz, equivalent to 2.9~$\rm{pb}^{-1}$ of integrated luminosity.

\begin{figure}
\centering
\includegraphics[width=0.49\linewidth,clip]{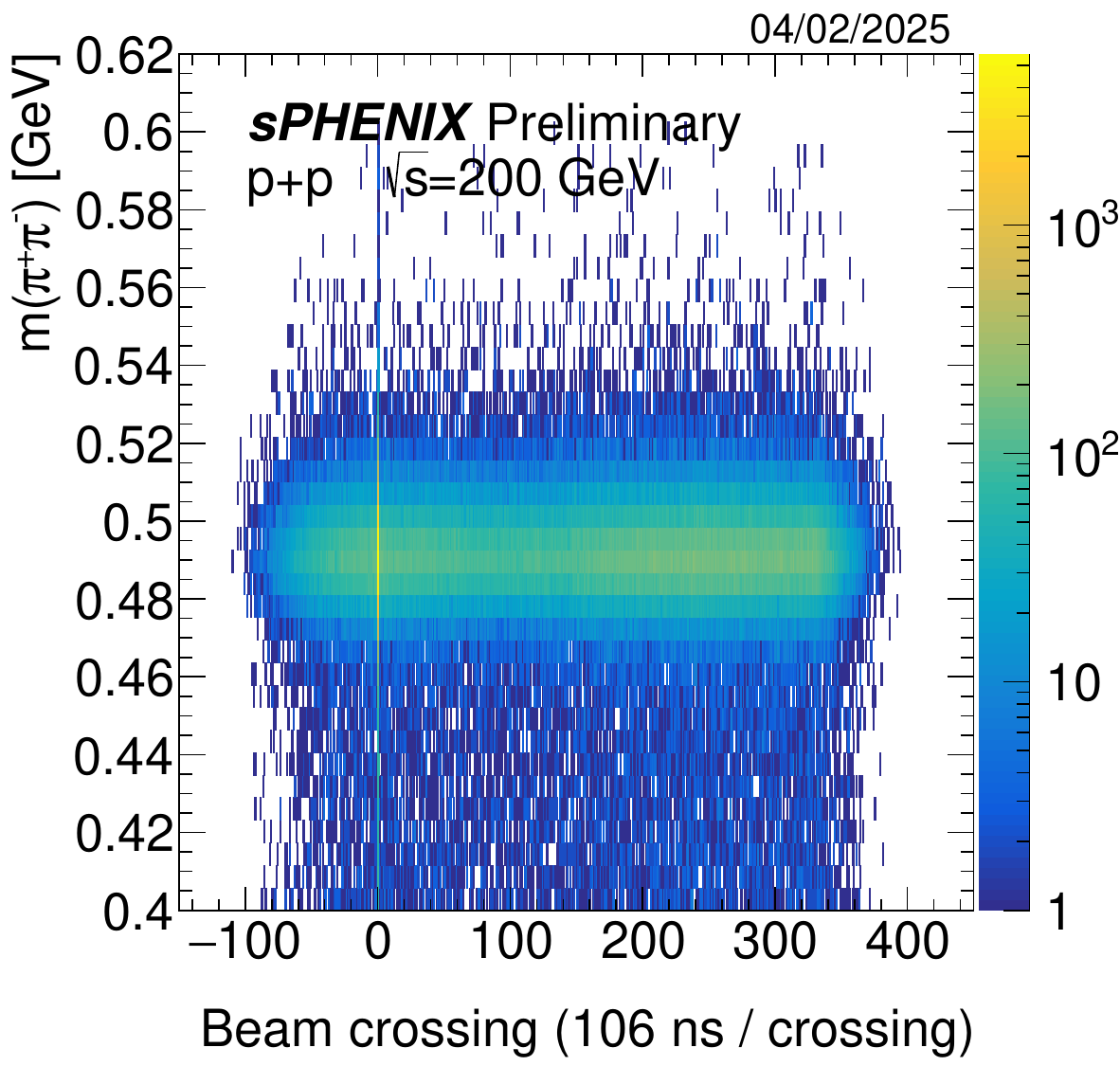}
\includegraphics[width=0.49\linewidth,clip]{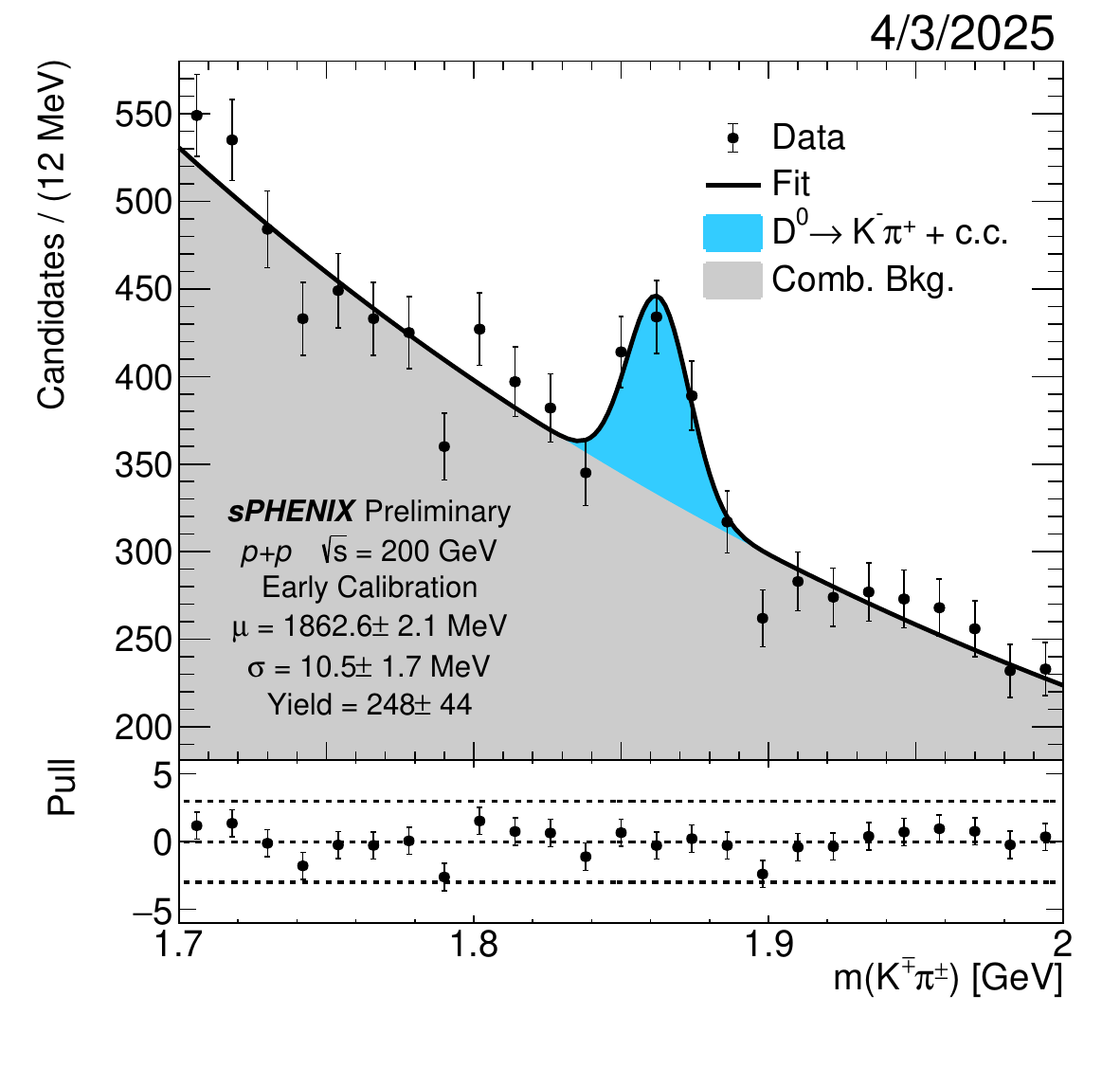}
\caption{(left) The invariant mass of reconstructed $K_s^0\rightarrow\pi^+\pi^-$ decays as a function of streaming readout beam crossing number in $\sqrt{s}=200$ GeV \pp collisions. (right) The invariant mass of reconstructed $D^0\rightarrow\pi K$ decays in $\sim$1 hour of data with initial calibrations.}
\label{fig:streamingkshort}       
\end{figure}

Tracks are reconstructed by seeding in the silicon and TPC, and then matching the track seeds in space and time before fitting with the Kalman Filter available in the Acts package~\cite{Osborn:2021zlr,Ai:2021ghi}. Resonances are reconstructed utilizing the KFParticle package adapted to the sPHENIX software stack. Figure~\ref{fig:streamingkshort} (left) shows the invariant mass of reconstructed $K^0_s\rightarrow\pi^+\pi^-$ decays as a function of the beam crossing number relative to the hardware trigger, where a crossing of 0 is coincident with the hardware trigger and all other crossings are $K_s^0$ from streamed data. An approximately constant efficiency of reconstructed $K_s^0$ is observed, indicating the streaming readout data acquisition and reconstruction was successful.

\section{Heavy Flavor Reconstruction}
\begin{figure}
\centering
\includegraphics[width=0.49\linewidth,clip]{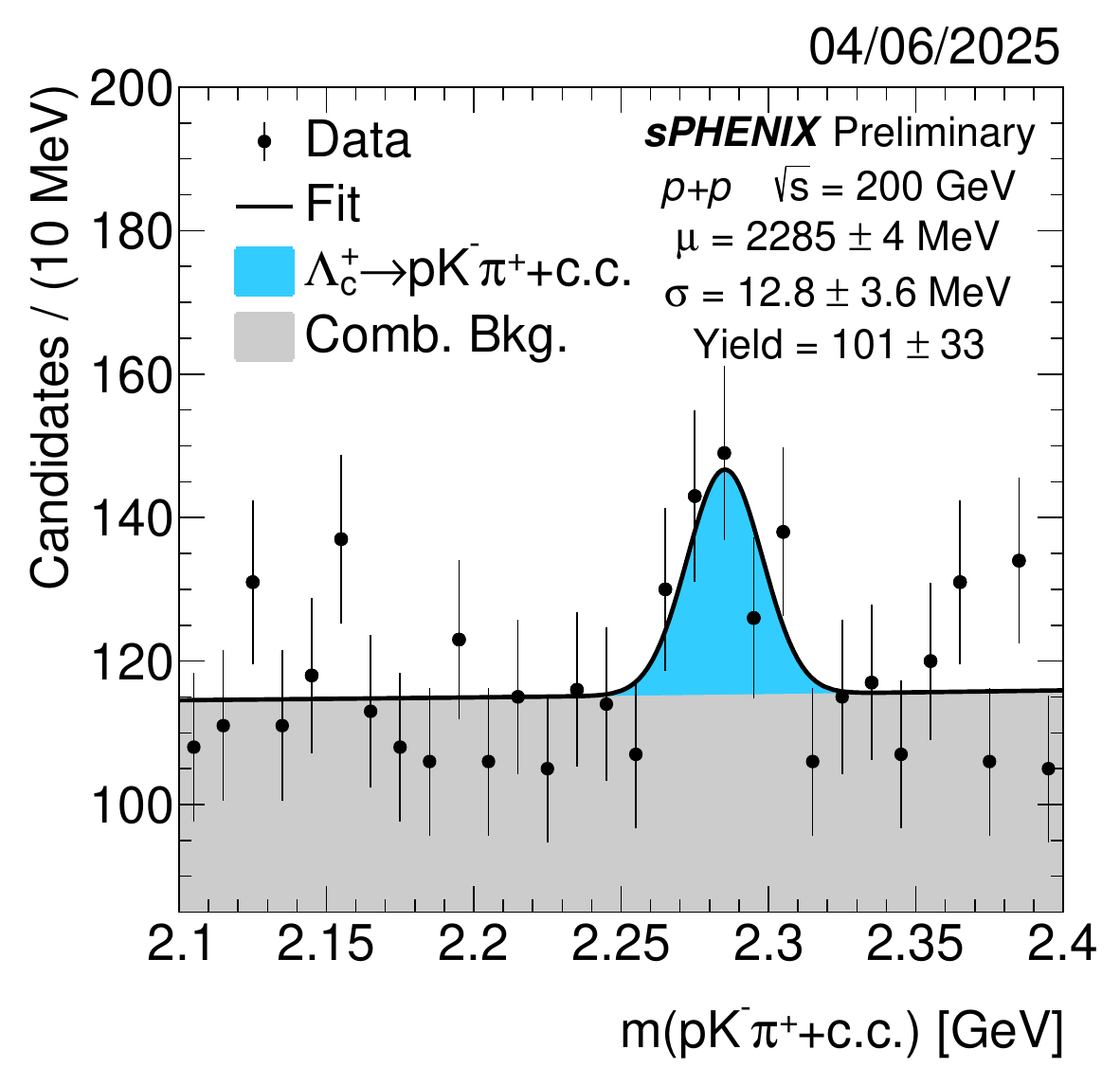}
\includegraphics[width=0.49\linewidth,clip]{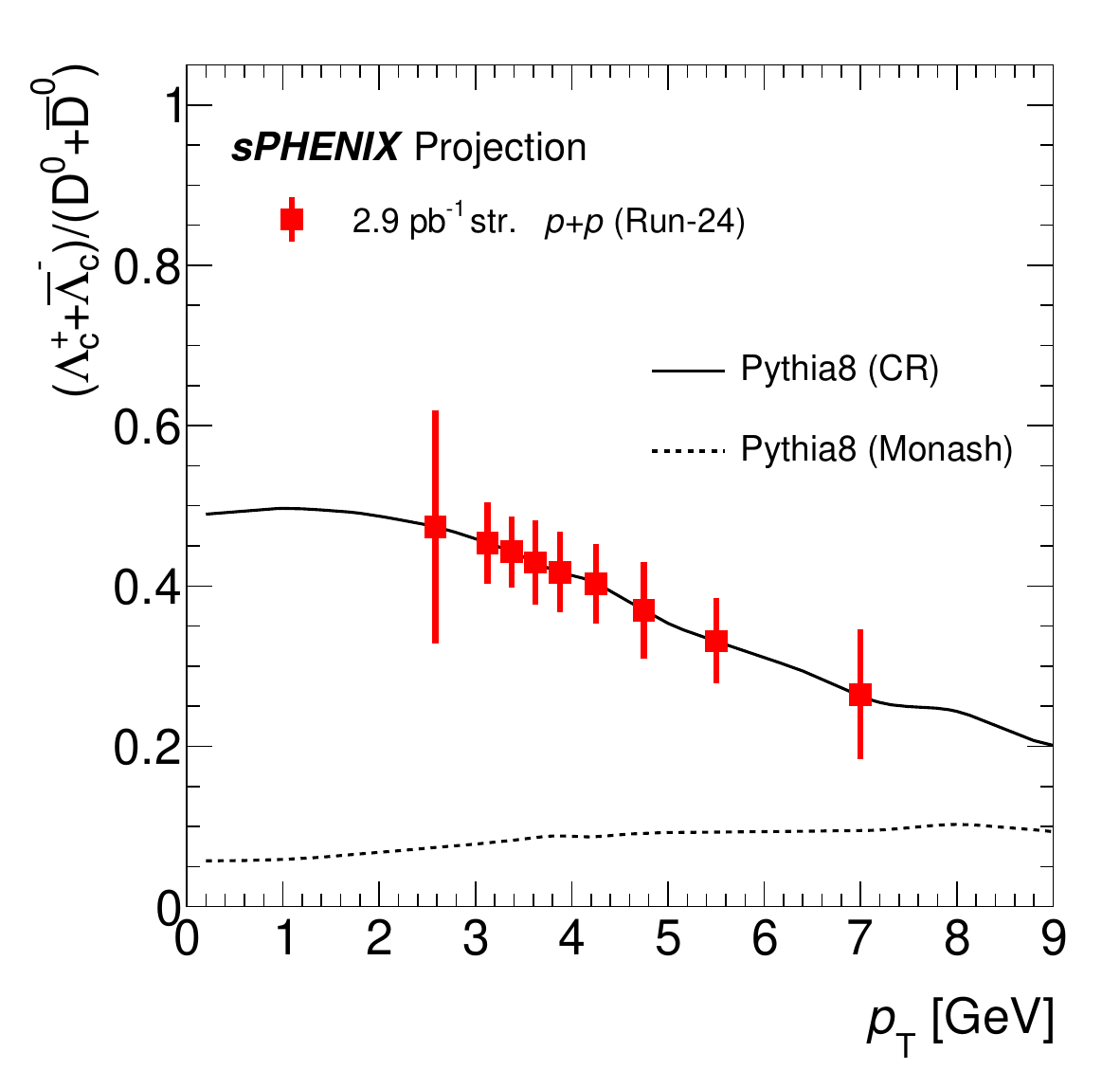}
\caption{(left) The first measurement of the \lambdac$\rightarrow pK\pi$ baryon in \pp collisions at RHIC energies at $3\sigma$ significance. (right) Projections for measurements of the \lambdac/\dzero ratio from the sPHENIX \pp data collected in 2024.
}
\label{fig:hf}       
\end{figure}

Analysis efforts for the first sPHENIX tracking data were focused on a small sample of the total integrated luminosity corresponding to approximately one hour of data. This was motivated by understanding and calibrating the first complete collision data measured by the sPHENIX tracking subsystems. The first initial calibrations are limited by the knowledge of the space charge distortions in the TPC, which dominates the momentum resolution of the tracks. Figure~\ref{fig:streamingkshort} (right) shows the first measurement of an invariant mass peak of the \dzero meson at approximately 5.6$\sigma$ significance from sPHENIX. Figure~\ref{fig:hf} (left) shows, for the first time in \pp collisions at RHIC, a measurement of the invariant mass peak of the \lambdac baryon at approximately 3$\sigma$ significance from sPHENIX. These show promising initial steps towards a complete measurement of the \lambdac/\dzero ratio, for which projections are shown in Figure~\ref{fig:hf} (right) based on the total integrated luminosity from the 2024 streaming readout data collection period.

A rich heavy flavor physics program at the sPHENIX experiment has begun with the first physics quality unbiased \pp data at $\sqrt{s}=$200 GeV data collected in 2024. The streaming tracking detectors and reconstruction have successfully demonstrated the measurement of a wide range of resonances~\cite{sPHENIXPub}, including the first heavy flavor resonances from sPHENIX and the first \lambdac measurement at RHIC in \pp collisions just six months after the first sPHENIX physics data taking campaign. sPHENIX is actively taking a large Au+Au dataset in 2025, with the goal of collecting 7.2~$\rm{nb}^{-1}$ of minimum bias data, that will enable a wide range of heavy flavor physics results providing precision measurements of the properties of the QGP.

%
\bibliography{bib}

\end{document}